# Optical Fiber Fault Detection and Localization in a Noisy OTDR Trace Based on Denoising Convolutional Autoencoder and Bidirectional Long Short-Term Memory

Khouloud Abdelli, Helmut Grießer, *Member, IEEE,* Carsten Tropschug, and Stephan Pachnicke, *Senior Member, IEEE*

*Abstract*—Optical time-domain reflectometry (OTDR) has been widely used for characterizing fiber optical links and for detecting and locating fiber faults. OTDR traces are prone to be distorted by different kinds of noise, causing blurring of the backscattered signals, and thereby leading to a misleading interpretation and a more cumbersome event detection task. To address this problem, a novel method combining a denoising convolutional autoencoder (DCAE) and a bidirectional long short-term memory (BiLSTM) is proposed, whereby the former is used for noise removal of OTDR signals and the latter for fault detection, localization, and diagnosis with the denoised signal as input. The proposed approach is applied to noisy OTDR signals of different levels of input SNR ranging from -5 dB to 15 dB. The experimental results demonstrate that: (i) the DCAE is efficient in denoising the OTDR traces and it outperforms other deep learning techniques and the conventional denoising methods; and (ii) the BiLSTM achieves a high detection and diagnostic accuracy of 96.7% with an improvement of 13.74% compared to the performance of the same model trained with noisy OTDR signals.

*Index Terms*— OTDR (optical time domain reflectometry), fiber optics, denoising, bidirectional long short-term memory, convolutional autoencoder, intelligent fault detection and diagnosis

## I. Introduction

FIBER monitoring has been commonly performed using optical time domain reflectometry (OTDR), an optoelectronic method widely used to characterize an optical fiber by exploiting the effects of Rayleigh scattering and Fresnel reflection [1]. The working principle of an OTDR is similar to that of a radar. The OTDR sends a series of light pulses into the fiber under test. Some of the light is scattered back towards the light source due to the Rayleigh scattering effect. The strength of the backscattered signal is recorded as a function of propagation time of the light pulse, which is converted then into the position on the optical fiber. Consequently, a characteristic OTDR trace, showing the positions of faults, including fiber misalignment/mismatch, fiber breaks, angular faults, dirt on connectors and macro-bends [2] along the fiber, is generated and used for event analysis. OTDR traces can be distorted by different sources of noise such as detector noise, electrical noise, thermal or shot noise. Such kinds of noise can interfere with the original OTDR signals which may cause deformation on the OTDR waveforms and thereby lead to inaccurate event detection and localization. Averaging multiple OTDR signals is usually a requisite to reduce the noise and thus to enhance the event detection capability. However, the averaging process is time consuming. As a fast measurement time is important to allow real-time fiber monitoring, it can be beneficial to denoise the OTDR signals before carrying out the event analysis and diagnosis.

Machine learning (ML) based approaches for OTDR event analysis have been proposed [3-7] to improve the event detection accuracy and to speed up the analysis. Trained with noisy OTDR data with SNR levels varying from 0 to 30 dB, compared to conventional techniques better detection and diagnostic capabilities were achieved [4,6,7]. However, these concepts perform poorly under low SNR values (SNR < 10 dB) and their generalization and robustness abilities underlying the capability and the effectiveness of the ML model in adapting and reacting to new unseen data may severely degrade further for SNR values lower than 0 dB. Lately, some techniques such as ensemble learning [8], structure improvement (e.g. probabilistic random forest [9]) and Bayesian neural networks [10] have been proposed to enhance the learning ability of the ML model in presence of noisy input data. But these novel approaches do not completely reduce the overfitting in order to prevent the model from memorizing the noise patterns. Therefore, the detection and diagnostic capability of the ML model under noisy data may not be improved a lot by only enhancing the robustness of the ML model to noise. That is why, an accurate and reliable event analysis requires the

This work has been performed in the framework of the CELTIC-NEXT project AI-NET-PROTECT (Project ID C2019/3-4), and it is partly funded by the German Federal Ministry of Education and Research (FKZ16KIS1279K).

K. Abdelli is with ADVA Optical Networking SE, Fraunhoferstr. 9a, 82152 Munich/Martinsried, Germany, and with Kiel University (CAU), Chair of Communications, Kaiserstr. 2, 24143 Kiel, Germany (e-mail: kabdelli@adva.com).

H. Grießer is with ADVA Optical Networking SE, Fraunhoferstr. 9a, 82152 Munich/Martinsried, Germany (e-mail: HGriesser@adva.com).

C. Tropschug is with ADVA Optical Networking SE, Märzenquelle 1-3, 98617 Meiningen, Germany (e-mail: CTropschug@adva.com).

S. Pachnicke is with Kiel University (CAU), Chair of Communications, Kaiserstr. 2, 24143 Kiel, Germany (e-mail: stephan.pachnicke@tf.uni-kiel.de).

removal of the noise from the input data, while keeping as many significant details of the desired signal as possible for the enhancement of the ML model's robust event detection and diagnostic capabilities. The latter requires to learn and extract the relevant information and features for solving the tasks of fault detection and diagnosis from a noisy input. Deep learning techniques have gained great popularity in the area of image denoising thanks to their good performance, but only very recently they have been applied in denoising one-dimensional signals. Specifically, ML models based on long short-term memory (LSTM) or convolutional neural networks (CNN) for denoising electrocardiogram (ECG) signals have been reported [11]. To the best of our knowledge, this is the first study applying and investigating ML methods for denoising OTDR signals.

In this paper, we propose a novel approach combining a denoising convolutional autoencoder (DCAE) and a bidirectional long-short term memory (BiLSTM) to solve the above problem. The DCAE is used to remove the noise from the OTDR signals before feeding the denoised signals to the BiLSTM model for event detection, localization, and diagnosis. The proposed method is validated by noisy experimental OTDR data with the SNR varying from -5 dB to 15 dB. Our contributions can be summarized as follows:

- an efficient denoising technique based on DCAE is proposed for reducing the noise from OTDR signals without causing loss in the signal information.
- a BiLSTM-based multi-task learning method for fault detection, localization, and diagnosis, trained with randomly corrupted noise-free OTDR signals, is proposed.
- The efficiency of the integrated learning approach combining DCAE and BiLSTM is validated using experimental monitoring data.

The rest of this paper is structured as follows: Section 2 gives some background information about the autoencoder, DCAE and BiLSTM. Section 3 presents the proposed DCAE and BiLSTM models as well as the combined framework. Section 4 describes the experimental setup and the validation of the presented combined approach. Conclusions are drawn in Section 5.

## II. BACKGROUND

### A. Denoising Convolutional Autoencoder

An autoencoder is a specific kind of an artificial neural network aiming to learn a compressed representation of an input in an unsupervised manner. It consists of an encoder and a decoder sub-model. The encoder is used to compress the input into lower-dimensional encoding (i.e., latent-space representation) and the decoder reconstructs the input from the compressed representation output of the encoder.

A denoising autoencoder (DAE) [12] is an extension and a stochastic version of the autoencoder. It reduces the risk of learning the identity function by randomly corrupting the input (e.g., by adding noise) and trying to reconstruct the original, uncorrupted input. Fig. 1 shows a standard architecture of the DAE. The input $x$ is corrupted by adding some noise to get $\tilde{x}$. Then, the encoder maps the noisy input $\tilde{x}$ to a low dimensional representation $z$ through a non-linear transformation, which is expressed as follows:

$$z = f(W\tilde{x} + b), \quad (1)$$

where $W$ and $b$ denote the weight matrix and the bias vector of the encoder respectively, and $f$ represents the activation function of the encoder.

The decoder reconstructs the output $\hat{x}$ given the representation $z$ via a nonlinear transformation, which it is given as follows:

$$\hat{x} = g(W'z + b'), \quad (2)$$

where $W'$ and $b'$ represent the weight matrix and the bias vector of the decoder respectively, and $g$ denotes the activation function of the decoder.

The DAE is trained to optimize the network parameters $\theta = \{W, b, W', b'\}$ by minimizing the reconstruction error between the output $\hat{x}$ and the input $x$, which is the loss function $L(\theta)$, typically the mean square error (MSE), defined as:

$$L(\theta) = \sum \|x - \hat{x}\|^2 \quad (3)$$

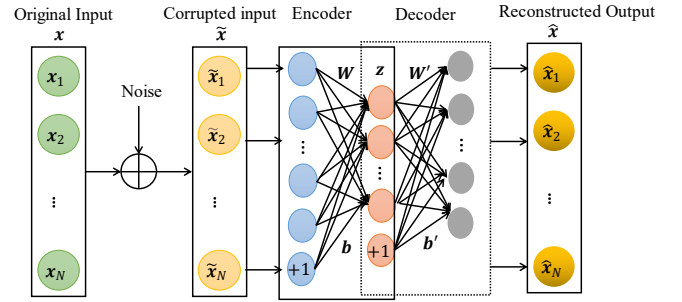

Fig. 1. Structure of a standard denoising autoencoder: the training objective is to minimize the reconstruction error between the output $\hat{x}$ and the original input $x$ (uncorrupted input).

A denoising convolutional autoencoder has the same standard DAE structure with convolutional encoding and decoding layers instead of fully connected layers. Each convolution layer consists of multiple kernels used to extract features or so-called feature maps. The latent representation $z_k$ of feature map $k$ is represented by:

$$z_k = f(W_k x + b_k) \quad (4)$$

The output of the decoder is expressed as:

$$\hat{x} = g\left(\sum_{k \in H} W'_k z_k + b'_k\right), \quad (5)$$

where H denotes the group of latent feature maps.

### B. Bidirectional Long-short Term Memory

LSTM [13] is a specific type of Recurrent Neural Network (RNN) used to process sequential data and to capture long-term sequential dependencies. The core computational unit of LSTM is called memory cell or block memory, containing weights and three gates, controlling the flow of information to the cell state. The forget gate decides what information to throw away from the cell state. The input gate determines what new information to store in the cell state, and the output gate decides what to output. As shown in Fig. 2, the previous cell state $c_{t-1}$ interacts with the previous cell output $h_{t-1}$ and the present input $x_t$ to determine, which elements of the internal state vector should be updated, kept, or discarded. The LSTM cell is updated by applying the following equations:

$$f_t = \sigma(W_{xf}x_t + W_{hf}h_{t-1} + b_f) \quad (6)$$
$$i_t = \sigma(W_{xi}x_t + W_{hi}h_{t-1} + b_i) \quad (7)$$
$$\tilde{c}_t = \tanh(W_{xc}x_t + W_{hc}h_{t-1} + b_c) \quad (8)$$
$$c_t = f_t \circ c_{t-1} + i_t \circ \tilde{c}_t \quad (9)$$
$$o_t = \sigma(W_{xo}x_t + W_{ho}h_{t-1} + b_o) \quad (10)$$
$$h_t = o_t \circ \tanh(c_t) \quad (11)$$

where $\sigma$ is the logistic sigmoid function, and $f$, $i$, $c$ and $o$ denote the forget gate, input gate, cell activation and output gate vectors, respectively. "∘" represents the Hadamard product operator, all $b$ are learned bias vectors, all $W$ are trainable weight matrices, and $\tilde{c}_t$ is a candidate cell value.

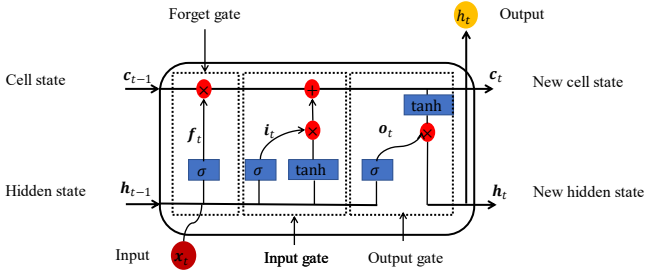

Fig. 2. Structure of the Long Short-Term Memory (LSTM) cell. The rectangles, the circles and the merging lines denote the gates, point-wise operations, and concatenation, respectively.

BiLSTM is an extension of LSTM that helps to improve the performance of the model. It consists of two LSTMs: one forward LSTM model that takes the input in a forward direction, and one backward LSTM model that learns the reversed input. The output $y_t$ of the model is generated by combining the forward output $\overrightarrow{h}_t$ and backward output $\overleftarrow{h}_t$ as described by the following equations.

$$\overrightarrow{h}_t = LSTM(x_t, \overrightarrow{h}_{t-1}) \quad (12)$$
$$\overleftarrow{h}_t = LSTM(x_t, \overleftarrow{h}_{t-1}) \quad (13)$$
$$y_t = W_{\overrightarrow{h}_t y}\overrightarrow{h}_t + W_{\overleftarrow{h}_t y}\overleftarrow{h}_t + b_y \quad (14)$$

III. PROPOSED APPROACH

A. Denoising Convolutional Autoencoder Model

The DCAE model for denoising OTDR signals is composed of an encoder and a decoder sub-model with 11 layers together, as shown in Fig. 3. The encoder takes as input a noisy OTDR sequence of length 100. It encodes the input into low dimensional features through a series of 5 convolutional layers containing 64, 32, 16, 64, and 32 filters (i.e., kernels) of size 16 × 1 with a stride (i.e., the step of the convolution operation) of 2, 2, 1, 1, 1, respectively. The stride is defined as the number of units or the amount by which the filter is slid over the input matrix at a time. Then the decoder attempts to reconstruct the output, given the compressed representation output of the encoder. The decoder is inversely symmetric to the encoder part. It consists of 6 transposed convolutional layers used to up-sample the feature maps. The last transposed convolutional layer with a single filter of size 16 × 1, and a stride of 1 is used to generate the output. Exponential linear units (ELU) are selected as an activation function for the hidden layers of the DCAE model, whereas for the output layer, there is no activation function. Each hidden layer is accompanied by batch normalization.

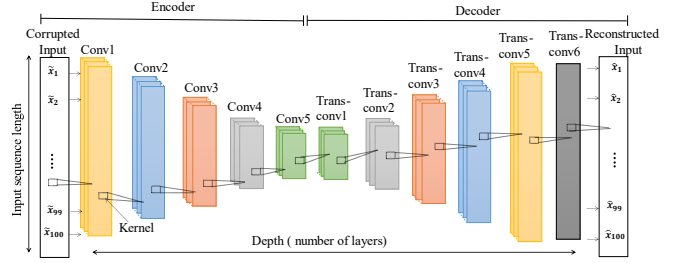

Fig. 3. Structure of the proposed DCAE model for OTDR signal denoising, Conv: convolutional layer, Trans-conv: transposed convolutional layer.

B. BiLSTM Model

A multi-task learning based BiLSTM model is proposed to simultaneously solve the following three tasks: fault detection $T_1$, fault localization $T_2$, and fault diagnosis $T_3$, given that the aforementioned tasks are highly related and can significantly benefit from the knowledge sharing across them in order to enhance the generalization capability of the model [4]. The model takes as input an OTDR sequence of length 100 and outputs concurrently the event type ($E_T$), the event position ($E_P$) and the event cause ($E_c$). As shown in Fig. 4, the architecture of the model is composed of a shared hidden layer consisting of one BiLSTM layer with 32 cells followed by three task-specific layers composed of 16, 20, and 16 neurons. BiLSTM is selected as a hidden shared layer as it is well suited to process OTDR sequential data and to capture long-term dependency. The model is trained by minimizing the loss function formulated as:

$$L_{total} = \sum_{i=1}^{3} \lambda_i \, l_{T_i}, \quad (15)$$

where $l_{T_i}$ denotes the loss of task $T_i$ and $\lambda_i$ represents the loss weight.

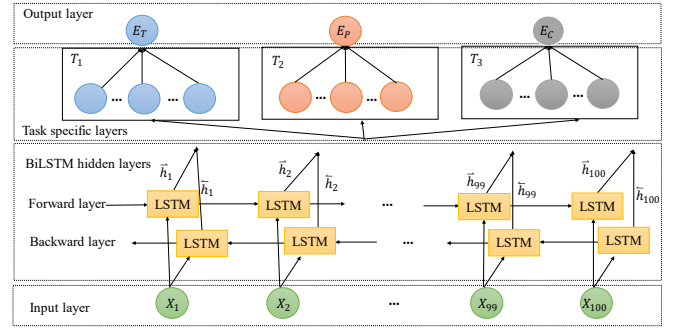

Fig. 4. Architecture of the proposed multi-task learning based BiLSTM model for fault detection, localization and diagnosis.

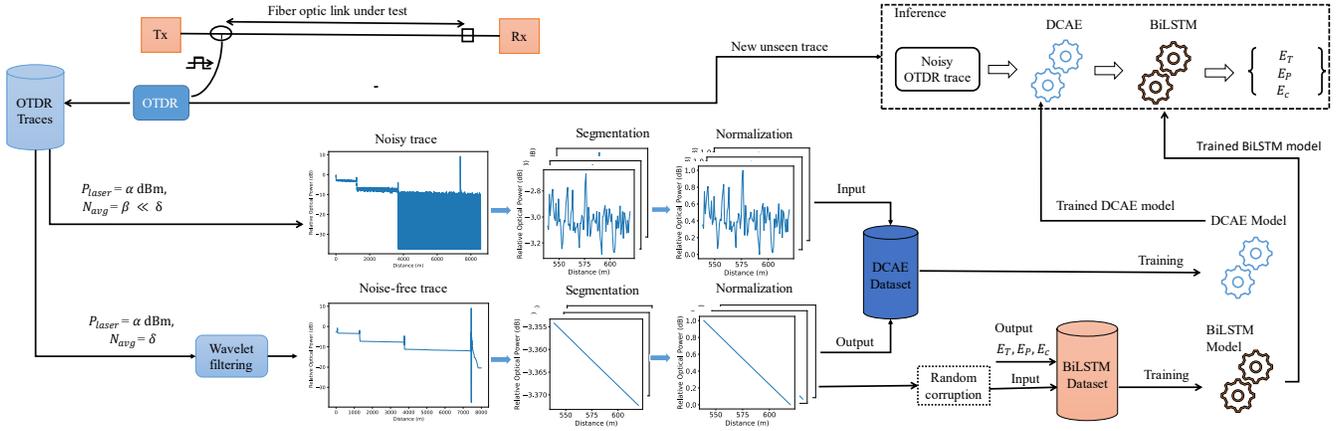

Fig. 5. Process for training and testing the combination of DCAE and BiLSTM models.

### C. Proposed Framework

Fig. 5 shows the proposed framework, which can be mainly split into four steps: (1) network monitoring and data collection, (2) dataset construction, (3) training of DCAE and BiLSTM models, (4) combining both models, DCAE and BiLSTM, and testing the complete framework using unseen noisy OTDR signals (inference). The fiber optic link is monitored using OTDR by performing averaging of repeated measurements. For each OTDR setting configuration (pulse width, laser power $P_{laser}$), a noise-free trace is generated by firstly averaging a large number of $\delta$ OTDR signals in order to get a signal with high SNR and then by performing a wavelet filtering on the trace, whereas a noisy trace is the result of averaging a very low amount of $\beta$ measurements to get a signal with an SNR level in the range of -5 dB to 15 dB. Given that the noise is white, it can not be indefinitely reduced by conducting a large number of averaging. Furthermore, some parts of the noise depend on the signal level, and there are some periodically repeating "wave noise parts" originating from the electronic parts and from the sampling, they are not really decreasing by performing a high averaging. Therefore, a wavelet filtering method is applied to the largely averaged OTDR signals to produce the "noiseless" traces by using the noise as an input for the wavelet filtering technique. Wavelet filtering, a time-frequency analysis technique based on signal decomposition into different frequency bands, has proven to be more effective in noise removal than other denoising methods such as Fourier analysis thanks to its multiresolution analysis characteristics. For our approach to generate noise-free signals, we adopted a total variation wavelet-based method producing nearly artifact free signal denoising. The details of the implementation of the used method are given in [14]. The noisy OTDR traces and their corresponding noise-free traces are segmented into sequences of length 100 and then normalized. After that, the dataset required for training the DCAE model is constructed by combining the noisy sequences (i.e., the input of the DCAE model) and the corresponding noise-free sequences (i.e., the desired output of the DCAE model). For BiLSTM model training, a dataset composed of randomly corrupted noise-free sequences (i.e., the input of the model) and their event characteristics, namely $E_T, E_P$ and $E_C$, labeled as the output of the model, is built. The random corruption of the noise-free sequences is performed by randomly selecting some values of the sequence equal to zero in order to enhance the robustness and improve the generalization capability of the model [10]. Then the DCAE is used to denoise the OTDR traces and the denoised signals are fed to the BiLSTM model to output the fault characteristics. The proposed approach is tested on unseen OTDR sequences with different SNR values ranging from -5 dB to 15 dB to evaluate its effectiveness.

### IV. VALIDATION OF THE PROPOSED FRAMEWORK

#### A. Experimental Setup

To validate the proposed approach, the experimental setup shown in Fig. 6 is employed. The setup is used to record noisy and noise-free OTDR traces incorporating different types of fiber faults. Optical components like connectors and variable optical attenuators (VOAs) are utilized to model specific events in the fiber optic link. The reflective faults (Fresnel reflection) with a sharp peak and an abrupt decrease are modeled by putting an open physical contact (PC) connector at the end of the setup (i.e., the end of the 5-km fiber). They are classified as perpendicular fiber-cuts. The non-reflective faults involving only small attenuation and no reflection are generated by placing an open angled physical contact (APC) connector at the end of the 1-km fiber. The coupler, VOA and the fibers are connected via fiber pigtails with APC connectors. Some of the connectors are deliberately not cleaned. As a result, merged events comprising overlapped non-reflective and reflective faults are caused. The attenuation of the non-reflective and combined faults is modified by varying the VOA settings. The OTDR configuration parameters, namely the pulse width, the wavelength and the sampling rate are set to 50 ns, 1650 nm and 8 ns, respectively. The laser power is varied from 7 dBm to 17 dBm. From 62 up to 320,000 OTDR records are collected and averaged.

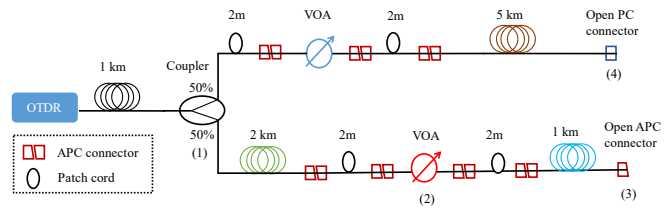

Fig. 6. Experimental setup for recording the OTDR signals.

#### B. Data Preprocessing

The generated noisy OTDR traces, with SNR values varying from -5 dB to 15 dB, and their corresponding noise-free traces are segmented into sequences of length 100 and normalized. For each noise-free sequence, the type $E_T$ (no event, reflective, non-reflective, merged), the position $E_P$ (defined as the index within the sequence), and the class $E_c$ (no event, fiber cut, fiber bend, dirty connector) are assigned. Two datasets; the first dataset containing the noisy and the noise-free sequences, as well as the second one incorporating the randomly corrupted noise-free sequences and the event characteristics are built for training the DCAE and BiLSTM models respectively. Each dataset, composed of 945,172 samples, is split into a training (60%), a validation (20%) and a test dataset (20%).

*C. Validation of the Denoising Capability*

*1. Overall performance of DCAE*

To assess the performance of the DCAE model, the reconstruction error (mean square error, MSE) and the SNR of the denoised sequence ($SNR_{out}$) are used as evaluation metrics. The reconstruction error is defined as the difference between the noise-free sequence and the denoised sequence. $SNR_{out}$ is calculated as follows:

$$SNR_{out} = 10 \times log_{10}\left(\frac{\sum_{i=1}^{i=N} x^2_i}{\sum_{i=1}^{i=N}(\hat{x}_i - x_i)^2}\right) \quad (16)$$

where $x_i, \hat{x}_i$ denote the values of the sampling point $i$ in the noise-free and the denoised sequences, respectively, and $N$ represents the length of the OTDR sequence.

The SNR improvement ($SNR_{imp}$) quantifies the difference between the SNR after noise reduction and the original noisy input sequence SNR:

$$SNR_{imp} = SNR_{out} - SNR_{in}, \quad (17)$$

where $SNR_{in}$ is obtained by the following expression:

$$SNR_{in} = 10 \times log_{10}\left(\frac{\sum_{i=1}^{i=N} x^2_i}{\sum_{i=1}^{i=N}(\tilde{x}_i - x_i)^2}\right) \quad (18)$$

and $\tilde{x}_i$ denotes the value of the sampling point $i$ of the noisy sequence.

Note that $SNR_{in}$ represents the computed SNR value of the noisy sequence and not the SNR of the whole trace as the SNR varies with changes of the signal along the trace.

Fig. 7 shows that the reconstruction error is very small (less than 0.0035) under all SNR conditions, which proves that DCAE is efficient in noise removal of OTDR signals thanks to its deep architecture. The results depicted in Fig. 7 demonstrate that the DCAE achieves significant $SNR_{imp}$ particularly for very low SNR levels. For example, for an SNR of -5 dB, the $SNR_{imp}$ can reach up to 21 dB. As the SNR increases, the $SNR_{imp}$ decreases.

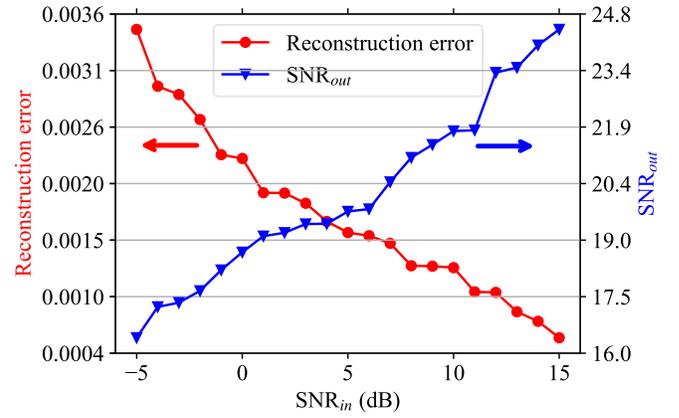

Fig. 7. Evaluation of DCAE performance using the reconstruction error and the output SNR of the denoised signal as metrics under different input SNR conditions.

To visually evaluate the performance of DCAE, a randomly unseen noisy trace is selected and the proposed denoising method is applied. Fig. 8 shows that the trace denoised by DCAE is very close to the noise-free trace. This demonstrates the effectiveness of DCAE in noise reduction, which can be very helpful to improve the event analysis.

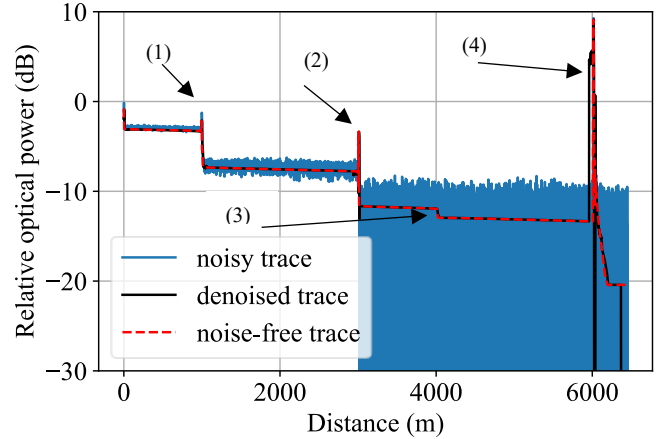

Fig. 8. Denoising of a random noisy trace using DCAE. The numbers denote the corresponding components shown in Fig. 6, inducing the faults.

To investigate the impact of the denoising on the spatial resolution measured at the fullwidth half maximum of a reflection event, we consider a noisy OTDR sequence incorporating a reflection peak induced by a reflector, and we denoise it using DCAE. Fig. 9 shows that the spatial resolutions of the noisy and denoised sequences are both 5 m, which proves that the denoising preserves the spatial resolution and does not deteriorate it.

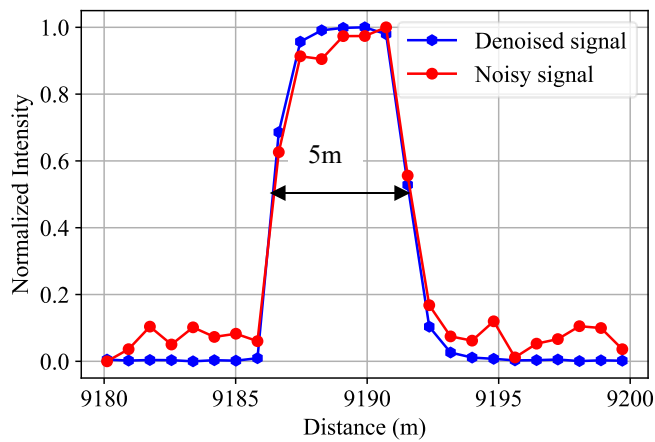

Fig. 9. Comparison of the spatial resolution without and with denoising.

*2. Comparison of DCAE with other ML models*

The DCAE model is compared with other ML techniques namely LSTM, DAE and CNN, using as evaluation metrics root mean square error ($RMSE$), percentage root mean square difference ($PRD$), and $SNR_{out}$.

The $RMSE$ is used for estimating the variance between the denoised signal (the output of the DCAE) and the noise-free signal (the actual output). It is expressed as:

$$RMSE = \sqrt{\frac{\sum_{i=1}^{i=N}(x_i - \hat{x}_i)^2}{N}} \quad (19)$$

The $PRD$ is used to evaluate the quality of the reconstructed signal. It is defined as:

$$PRD = \sqrt{\frac{\sum_{i=1}^{i=N}(x_i - \hat{x}_i)^2}{\sum_{i=1}^{i=N} x_i^2}} \times 100 \quad (20)$$

Please note that a better denoising technique should have lower $RMSE$ and $PRD$ and a higher $SNR_{out}$.

Figs. 10 (a) and (b) illustrate the average RMSE and PRD scores for input SNR levels ranging from -5 dB to 15 dB. For the different evaluated ML models, as SNR increases, the RMSE and PRD decrease. It is observed that DCAE outperforms the other ML methods as it achieves the lowest RMSE and PRD scores for all input SNR conditions. Although that for high SNR values (SNR > 10 dB), the RMSE and PRD scores obtained by each ML model are getting close, DCAE still achieves smaller RMSE, and PRD compared to the other ML techniques. The results demonstrate that DCAE is more effective in noise removal compared to other ML approaches particularly at low SNR levels.

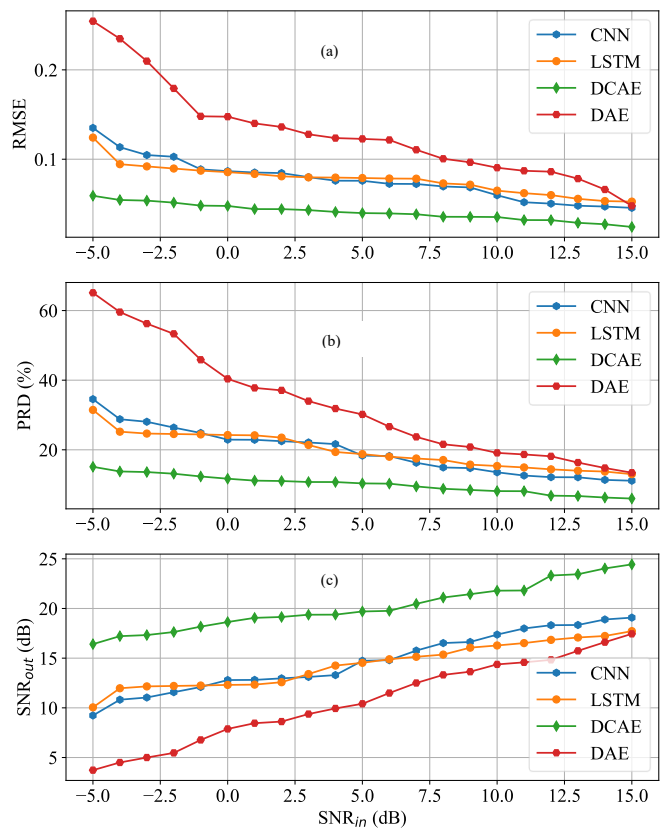

Fig. 10. Comparison of denoising performance of different ML methods under input SNR conditions ranging from -5 dB to 15 dB: (a) RMSE, (b) PRD, (c) $SNR_{out}$ of ML models at different SNR levels.

Fig. 10 (c) shows that the average $SNR_{out}$ scores for different SNRs vary from -5 dB to 15 dB. It can be seen that DCAE achieves better denoising performance compared to CNN, LSTM and DAE as it yields higher $SNR_{out}$ under all input SNRs. For low input SNR levels, the $SNR_{imp}$ is high, while when the input SNR increases, the $SNR_{imp}$ is getting lower.

As shown in Fig. 10, DCAE achieves better denoising than the other ML methods, nevertheless LSTM, DAE and CNN are also effective in noise removal. As depicted in Fig. 11, the traces denoised by the different ML approaches are close to the noise free trace.

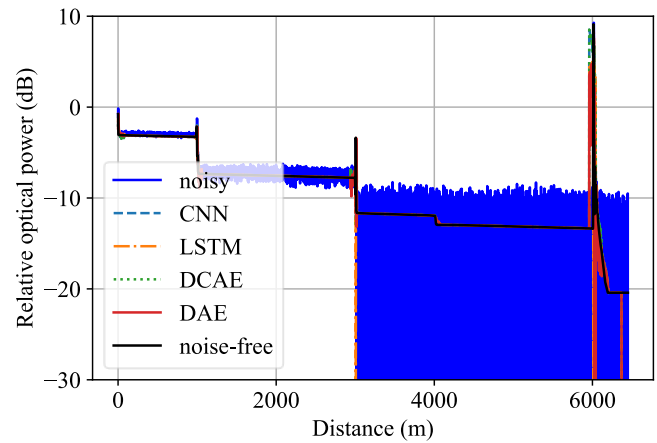

Fig. 11. Denoising of an example trace with different ML methods.

### 3. Comparison of DCAE with conventional methods

The DCAE model is compared with two conventional denoising techniques, namely low pass filter and wavelet denoising, again using as evaluation metrics RMSE and PRD. To achieve better performance, the parameters of the conventional methods, namely the cutoff frequency, the wavelet function (i.e., wavelet mother) and the number of decomposition levels, are optimized for each SNR level. The used filter is the third-order low pass Butterworth filter.

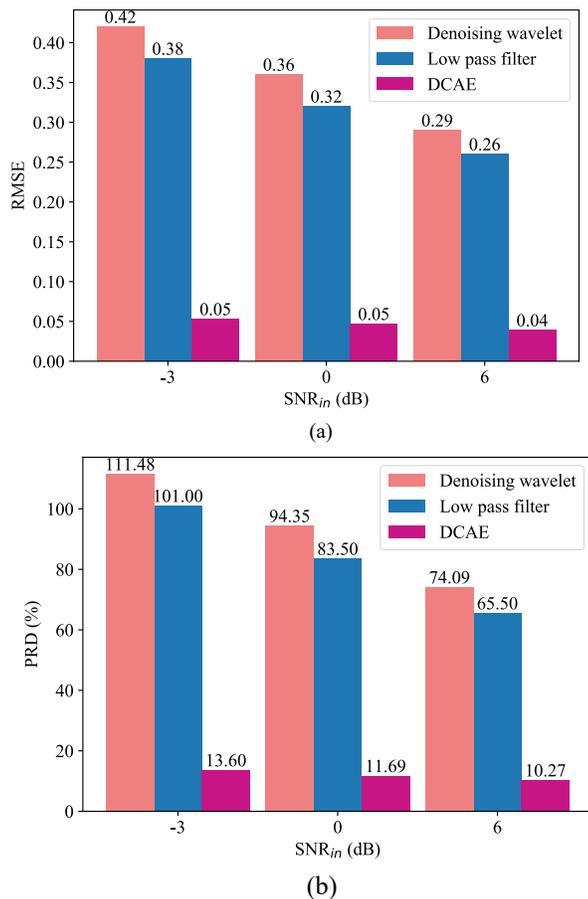

Fig. 12. Comparison of denoising performance of the DCAE model with conventional denoising methods under different input SNR conditions: (a) RMSE, (b) PRD of DCAE, filtering and wavelet denoising techniques with varying input SNRs.

The selected mother wavelet functions for the different SNR levels are either fourth order bi-orthogonal splines with 2 decomposition levels or second, third or seventh order symlets with decomposition levels of 4, 3 and 1, respectively. Fig. 12 shows the RMSE and PRD scores for different SNR levels. The results prove that DCAE outperforms the conventional denoising techniques by achieving the smallest values of RMSE and PRD for the different input SNRs. For an input SNR of -3 dB, DCAE yields an RMSE of 0.05 with a PRD of 13.6%, while low pass filter and denoising wavelets achieve RMSE values of 0.38 and 0.42 as well as PRD values of 101% and 111.48%, respectively.

### 4. Optimization of DCAE

The depth of the model, the kernel size and the length of the input sequence have a significant influence on the performance of DCAE. As depicted in Fig. 13, the reconstruction error of DCAE shows a decreasing trend with the increase of the depth for a different number of filters per convolutional layer, before reaching the depth of 11. Increasing the depth helps the DCAE model to learn a large number of parameters or features and thereby to learn more efficiently more complex representations of the different events. However, widening the layers too much (higher than 11) can lead to overfitting and thus reduces the performance of the DCAE. The kernel size influences as well the denoising effect of the model. The reconstruction error indicates a descending trend with an increase of the kernel size before reaching the optimum kernel size value of 16. A smaller kernel size can capture a lot of information which leads to overfitting, whereas a larger kernel size can cause the loss of the information leading to underfitting. The selection of the input sequence length impacts also the denoising capability of the model. For an input sequence of length higher than 100, the reconstruction errors under different SNR conditions increase with the sequence length. For an input sequence of length 50, the denoising capability is worse, which indicates that DCAE could not effectively learn the relevant features of the signal due to the small input sequence length leading to the loss of the information. As the length of the input sequence increases (higher than 50), more useful information mixed with noise information is given to the DCAE model. Therefore, the performance of DCAE for an input sequence length higher than

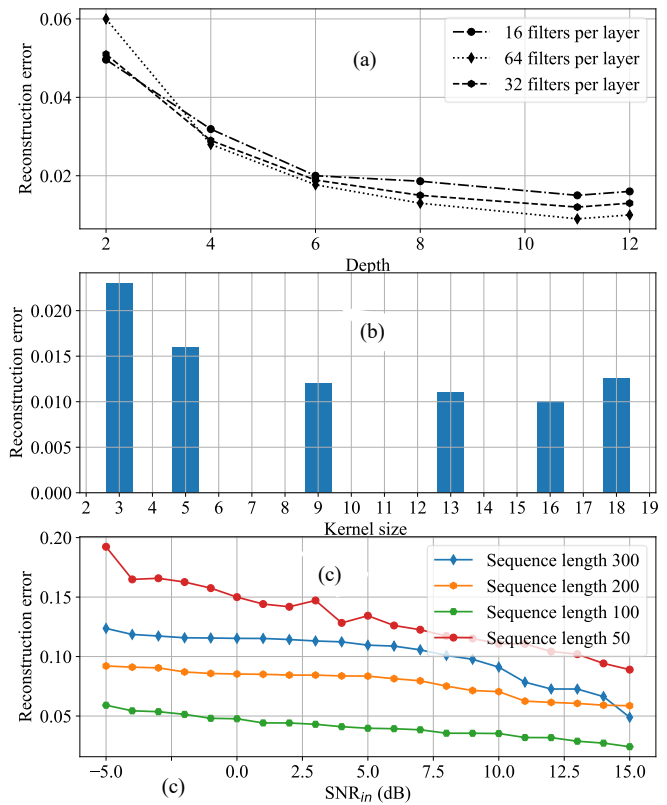

Fig. 13. Optimization of the DCAE model: (a) reconstruction errors with different depths for different filters per layer, (b) with various kernel sizes, (c) for different input sequence lengths.

50 is improved. However, as the input sequence's length is getting higher than 100, the performance of the DCAE is reduced as it is getting more challenging for DCAE to capture the relevant information underlying the event's shape given the large amount of no signal information predominant in the sequence and resulting in a partial loss of signal information. The results demonstrate that the input sequence of length 100 is the optimum.

*5. Robustness Investigation of DCAE*

The robustness of the DCAE model is investigated for the following three scenarios:

- We modify the experimental setup used to train DCAE by removing some components as shown in Fig. 14 (a) to generate different OTDR traces.
- We test the DCAE denoising capability on unseen OTDR traces recorded by a completely new experimental setup used for point-to-point link monitoring, as depicted in Fig. 14 (b), exploiting a different operating regime with a dedicated reflector and low sensitivity.
- We evaluate the DCAE performance in denoising OTDR traces for passive optical network monitoring, generated by a more complex experimental setup, which is shown in Fig. 14 (c) [15].

The OTDR traces recorded by the different experimental setups are fed to DCAE for denoising. Fig. 15 illustrates the denoised traces for the different scenarios. It can be seen that the DCAE is able to accurately reconstruct the noise-free signal from the noisy one while preserving the shape of the signal, which proves that the DCAE effectively learns the signal features of the OTDR trace and not the noise features.

### D. VALIDATION OF THE DETECTION AND DIAGNOSIS CAPABILITY

The proposed method of combining DCAE and BiLSTM, is compared to the BiLSTM model without applying any denoising technique, and trained with noisy OTDR sequences with SNR values varying from -5 dB to 15 dB. The results shown in Fig. 16 demonstrate that the combination of DCAE and BiLSTM achieves better event detection capability compared to the BiLSTM model without denoising, and that the denoising step performed by DCAE helps to significantly improve the detection accuracy particularly under very low SNR levels ($SNR \leq 0\ dB$). It can be seen that for an input SNR of -1 dB, the accuracy obtained by the combined method is 91.66%, whereas the accuracy of the BiLSTM model without denoising is only 76.92%, which proves that the proposed approach achieves higher event detection performance even for low SNRs.

The results of the comparison of the event localization estimation's accuracy of the "DCAE+BiLSTM" combination and BiLSTM without denoising depicted in Fig. 17, prove that the combined approach achieves smaller event position errors compared to BiLSTM without denoising. For example, for an SNR level of -4 dB, the combined method yields an event position error of 6 m, while the error obtained by BiLSTM without denoising is 11 m.

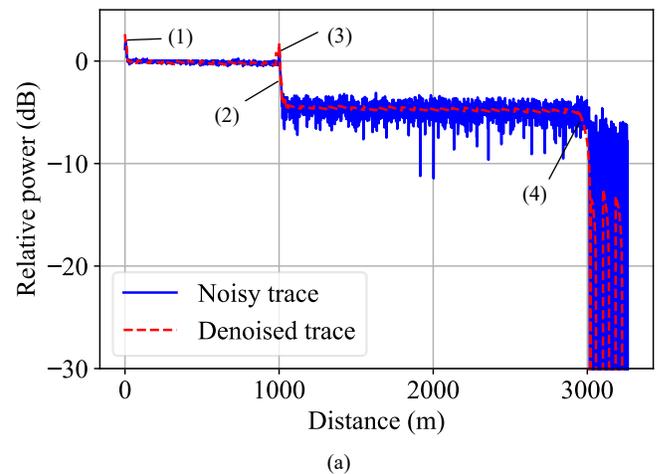

(a)

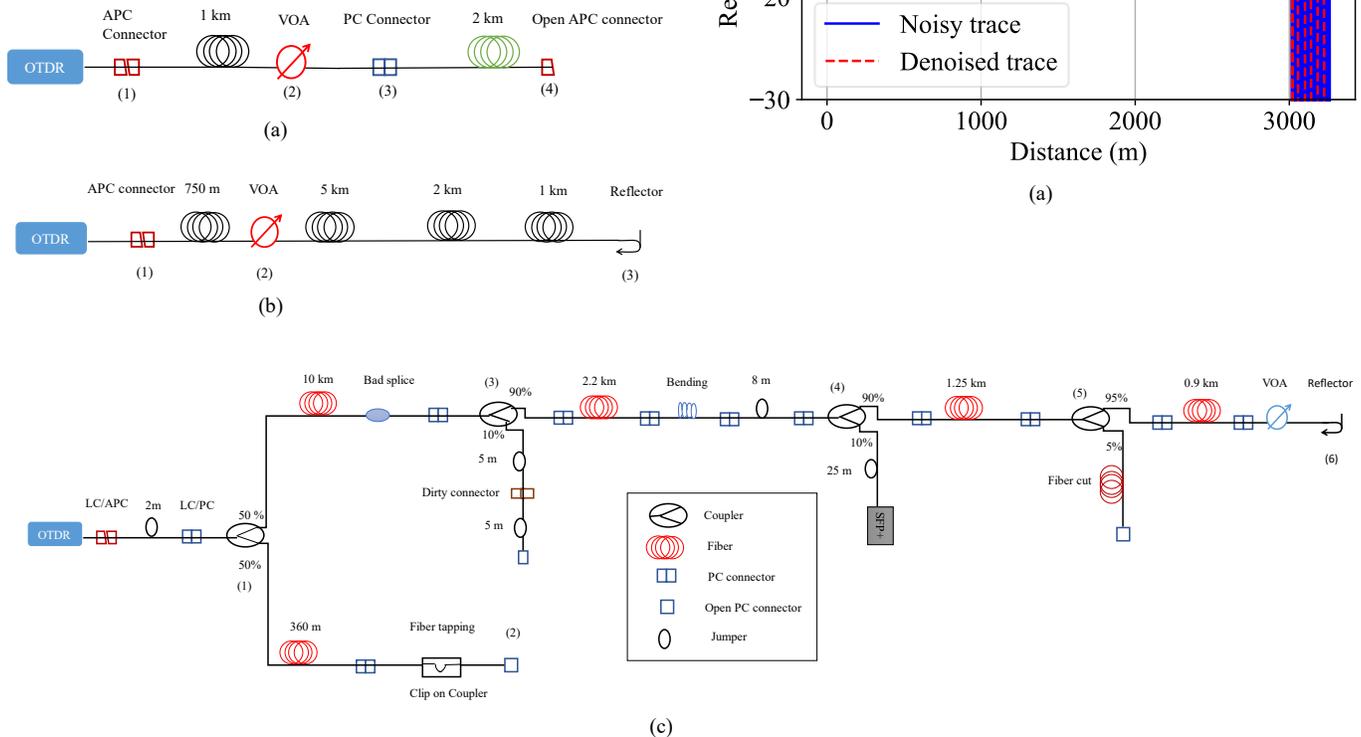

Fig. 14. Experimental setups for testing the robustness of DCAE.

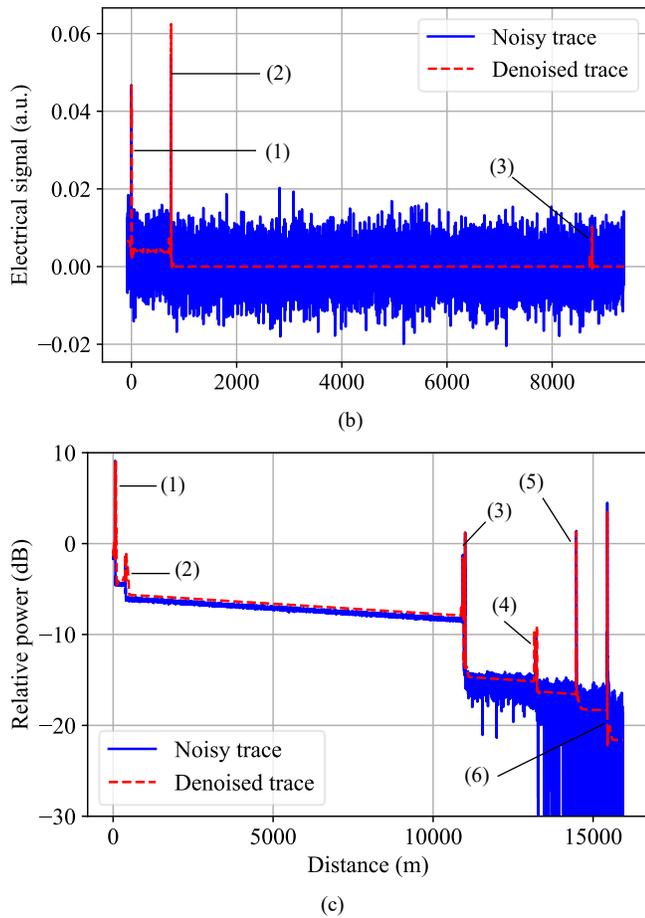

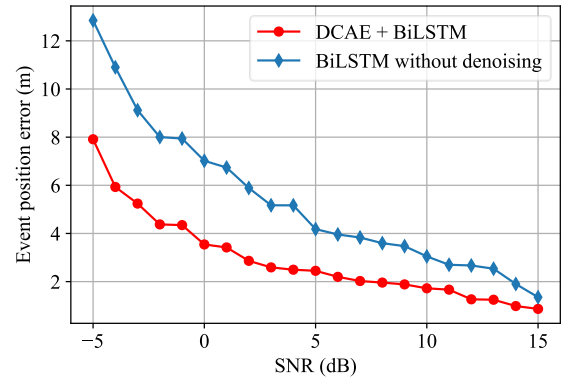

Fig. 17. Event localization error of BiLSTM model with and without denoising under SNR values varying from -5 dB to 15 dB.

It can be seen that the BiLSTM model without denoising highly misclassifies the fiber bend as no event since under low SNR conditions, it is trickly to distinguish the fiber bend's pattern from a normal signal's pattern due to the noise. Whereas the combined method achieves a better classification rate of a fiber bend, which proves that the denoising process of DCAE helps to significantly improve the classification rate of the fault's cause particularly for the fiber bend and no event types.

Fig. 15. Denoising results of DCAE for the different setups: (a) noisy trace generated by setup 14 (a) and the denoised trace by DCAE, (b) noisy trace recorded by setup 14 (b) and the denoised trace by DCAE, (c) noisy trace generated by setup 14 (c) and the denoised trace by DCAE, the number indicates the component causing the event.

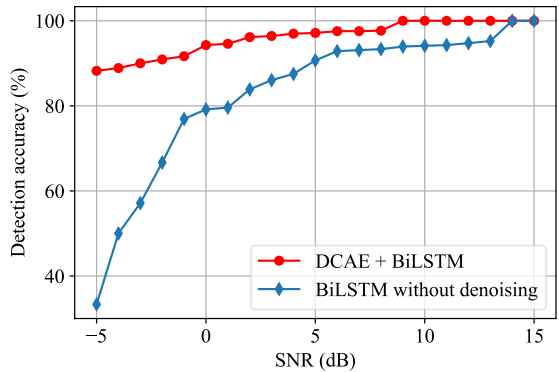

Fig. 16. Detection accuracy of BiLSTM model with and without denoising.

The results of the confusion matrices of the event diagnosis for the combined approach and BiLSTM without denoising under an input SNR condition of 0 dB, are illustrated in Fig. 18. The overall diagnosis accuracy of the "DCAE+BiLSTM" combination and the BiLSTM model without denoising are 96.37% and 80.13%, respectively.

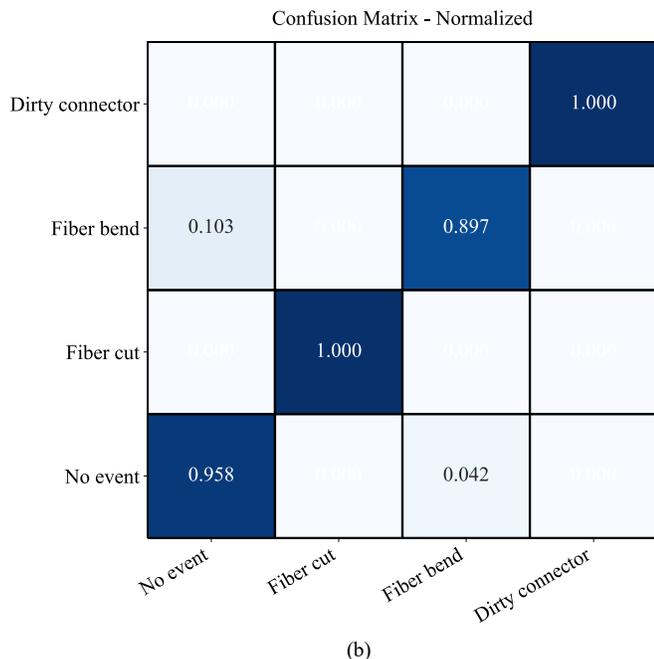

(b)

Fig. 18. Confusion matrices of diagnosis results under SNR condition of 0 dB: (a) the confusion matrix of BiLSTM without denoising, (b) the confusion matrix of DCAE + BiLSTM.

The feature learning ability of the proposed approach under very low SNR conditions ($SNR \leq 0\ dB$) for solving the task of event diagnosis is visually investigated using the t-distributed stochastic neighbor embedding (t-SNE) technique [16]. Fig. 19 shows that the input features of the different event causes overlap and are of poor separability. In contrast the learned features of the combined model are almost discriminative, which demonstrates that the proposed model can learn efficiently the relevant features characterizing each event type, leading to high event diagnosis accuracy under low SNRs. The overlap between the learned features of a few fiber bend cases, and of "no events" shows that the model might misclassify the fiber bend as "no event" for low SNR levels.

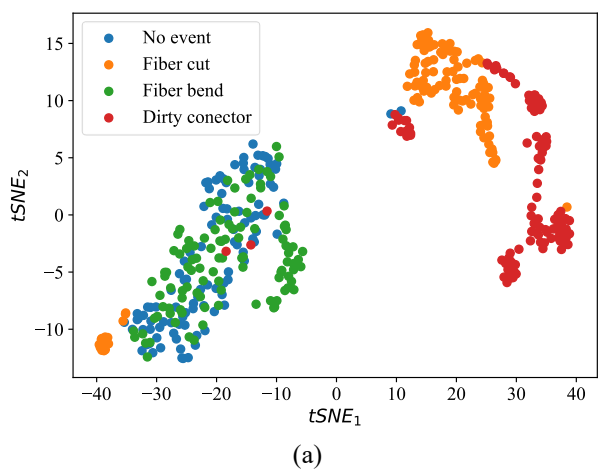

(a)

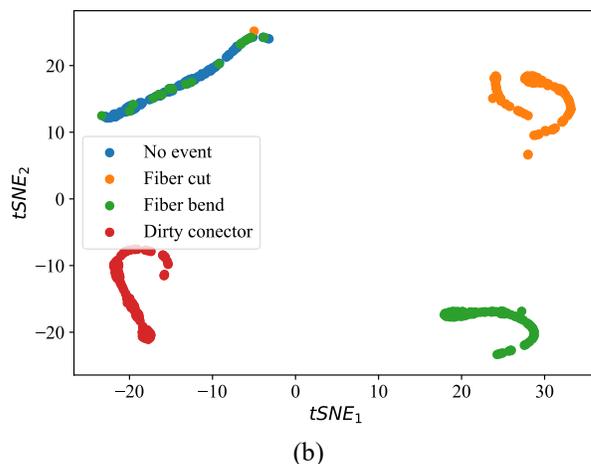

(b)

Fig. 19. Visualization of feature learning with low SNR: (a) the input features, (b) the learned features with the proposed approach.

Fig. 20 shows the output of the integrated learning approach, combining DCAE and BiLSTM models, for a noisy OTDR trace. The ML approach accurately detects and identifies the different events within the trace, and it achieves a low misclassification rate (two positions are misclassified as fiber bend). The predicted positions of the events are very close to the true location of the faults.

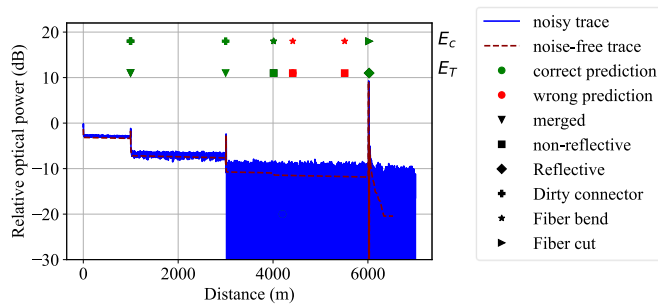

Fig. 20. Output of the ML method combining DCAE and BiLSTM, compared to the noisy OTDR trace.

V. CONCLUSION

A combined approach of a denoising convolutional autoencoder (DCAE) and a bidirectional long short-term memory (BiLSTM) is proposed to tackle the problem of optical fiber fault detection, localization and diagnosis under very low SNR conditions varying from -5 dB to 15 dB. The DCAE is used to denoise the OTDR signals, then the BiLSTM performs the event analysis with the denoised signals from the DCAE. The combined method is validated by experimental noisy OTDR signal datasets. The results prove that the DCAE is effective in noise reduction without impacting the signal information, since the reconstruction error is close to zero even for low SNRs. Moreover, the denoising capability of DCAE is much better than with other ML techniques, namely a denoising autoencoder, a convolutional neural network and an LSTM, particularly under very low SNR conditions ($SNR \leq 0\ dB$). As a third point the DCAE model outperforms the conventional denoising techniques; low pass filter and wavelet denoising. Finally, with the denoising process performed by DCAE, the BiLSTM model achieves high event detection, event localization and event diagnosis accuracy at very low SNR levels: the average detection accuracy of the BiLSTM under SNR conditions lower than 0 dB is 90.66%, whereas without denoising, the accuracy is 60.53%, which demonstrates the importance of the denoising step to achieve accurate fault analysis under noisy SNR conditions.